\documentclass[12pt]{iopart}

\usepackage{graphicx}
\usepackage{iopams}
\begin{document}

\title{Sub-electron Charge Relaxation via 2D Hopping Conductors}

\author{Yusuf A. Kinkhabwala\dag and Konstantin K. Likharev\dag}
\address{\dag\ Department of Physics and Astronomy, Stony Brook
University, Stony Brook, NY 11794-3800}

\begin{abstract}
We have extended Monte Carlo simulations of hopping transport in
completely disordered 2D conductors to the process of external
charge relaxation. In this situation,  a conductor of area $L \times
W$ shunts an external capacitor $C$ with initial charge $Q_i$. At
low temperatures, the charge relaxation process stops at some
``residual" charge value corresponding to the effective threshold of
the Coulomb blockade of hopping. We have calculated the r.m.s$.$
value $Q_R$ of the residual charge for a statistical ensemble of
capacitor-shunting conductors with random distribution of localized
sites in space and energy and random $Q_i$, as a function of
macroscopic parameters of the system. Rather unexpectedly, $Q_{R}$
has turned out to depend only on some parameter combination: $X_0
\equiv L W \nu_0 e^2/C$ for negligible Coulomb interaction and
$X_{\chi} \equiv LW \kappa^2/C^{2}$ for substantial interaction.
(Here $\nu_0$ is the seed density of localized states, while
$\kappa$ is the dielectric constant.) For sufficiently large
conductors, both functions $Q_{R}/e =F(X)$ follow the power law
$F(X)=DX^{-\beta}$, but with different exponents: $\beta = 0.41 \pm
0.01$ for negligible and $\beta = 0.28 \pm 0.01$ for significant
Coulomb interaction. We have been able to derive this law
analytically for the former (most practical) case, and also explain
the scaling (but not the exact value of the exponent) for the latter
case. In conclusion, we discuss possible applications of the
sub-electron charge transfer for ``grounding" random background
charge in single-electron devices.
\end{abstract}


\maketitle

\section{\label{sec:level1} Introduction}

Electron transport via inelastic hops between localized states in
disordered conductors has been studied for many years, with the main
focus on the average transport characteristics (e.g., dc current
dependence on temperature and applied electric field) and to a
lesser extent on the $1/f$ noise - see
Refs.~\cite{MottBook,ShklovskiiBook,EfrosPollackCollection,KoganBook}
for comprehensive reviews of this work. The relatively recent
observation \cite{AverinLikharev1991, MatsuokaLikharev1998,
Review2002} that hopping transport may provide quasi-continuous
(``sub-electron") charge transfer gave a motivation for the
extension of this work to the statistics of the electric charge $Q$
carried over by the hopping current.

The idea of the quasi-continuous charge transfer is quite simple:
due to the electrostatic polarization, each electron hop between two
localized sites inside the conductor leads to a step-like increase
of the ``external charge" $Q(t)$, which may be defined as the time
integral of current $I(t)$ flowing through the wires connecting the
conductor's electrodes to the electric field source. If an electron
is transferred through the whole sample in one hop (as happens in
the usual tunnel junctions), the charge step $\vert \Delta Q \vert$
is equal to the fundamental charge $e$. However, if an electron in
an extended conductor hops between two sites which are separated by
a distance $\Delta r$ much less than the conductor length $L$, then
the step $\vert \Delta Q \vert$ is of the order of $e \times (\vert
\Delta r \vert /L) \ll e$. (The exact expression depends on the
sample and electrode geometry.) This means that the charge transport
becomes nearly continuous, just as in long diffusive conductors
\cite{Review2002,NavehAverinLikharev1998}. This phenomenon may have
several useful applications in single-electronics, especially since
the hopping conductors (in contrast to their diffusive counterparts)
may provide the necessary high values of resistance $R \gg \hbar
/e^2$ \cite{Likharev1999} without adding too much stray capacitance
to that of single-electron islands.

One of the manifestations of the quasi-continuous charge transport
is the suppression of the shot noise \cite{KoganBook,
JongBeenakker1997, BlanterButtiker2000}. Namely, for sufficiently
small values of the observation frequency $f$ (with a possible
exception for the $1/f$ noise at very low frequencies) the current
noise spectral density $S_{I} \left( f \right)$ becomes
approximately $L_c/L \ll 1$ times the Schottky value $2eI$, where
$L_c$ is some characteristic length scale. This prediction
\cite{AverinLikharev1991} has been confirmed in several recent
experimental \cite{Kuznetsovetal2000,Roshkoetal2002} and theoretical
\cite{1DKorotkovLikharev2000, 2DSverdlovKorotkovLikharev2001,
2DCLP-KinkhabwalaSverdlovKorotkovLikharev2004,
2DCIP-KinkhabwalaSverdlovKorotkovLikharev2004} studies of hopping.

\begin{figure}
\begin{center}
\includegraphics[height=2.6in]{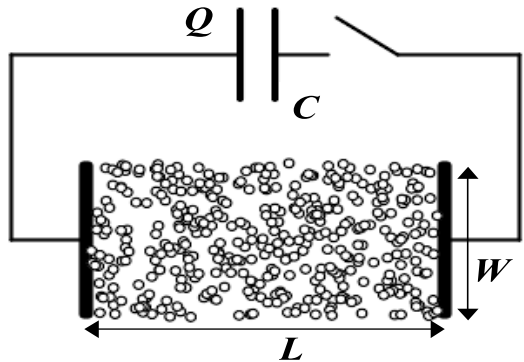}
\end{center}
\caption{The system under analysis (schematically).}
\label{fig:deviceschematic}
\end{figure}

The goal of this work has been to study another manifestation of the
quasi-continuous charge transfer at hopping, which is more closely
related to its most important potential application: the ability to
``ground" sub-electron amounts of electric charge
\cite{Likharev1999}. For this, we have analyzed the simple system
shown in Fig.~\ref{fig:deviceschematic}: a hopping conductor shunts
an external capacitance $C$ with an initial charge $Q_i$. The
capacitance charge $Q$ leads to a nonvanishing electric field
$E=V/L=Q/CL$ applied to the conductor, which causes electrons to hop
through the conductor. These hops result in the gradual reduction of
the charge $Q$ and hence the field $E$. At the perfectly continuous
(``Ohmic") conduction the process would continue until $Q$ and $E$
vanished completely (at $T \rightarrow 0$); however, for hopping
conductors of a finite size $L \times W$ the charge relaxation stops
at a certain finite residual charge which generally depends not only
on the macroscopic parameters of the system, but also on the
particular distribution of the localized sites over space and energy
and on the initial charge $Q_i$.

Though qualitative experimental evidence of sub-electron charge
relaxation has been obtained long ago \cite{Lambe,Kuzmin}, to the
best of our knowledge this phenomenon has never been studied in
detail. The objective of this work has been to study the dynamics of
this charge relaxation process, and the statistics of the residual
charge theoretically. The problem is essentially classical, but
multi-particle, highly nonlinear, and statistical, so that most
results have to be obtained by numerical (Monte Carlo) simulation
using modern supercomputer facilities (see the Acknowledgments
section below).

\section{\label{sec:level1} Model}

For the hopping conductor, we have used the same model whose average
transport characteristics and current noise had been extensively
explored recently for the case of fixed, constant applied field $E$
\cite{2DCLP-KinkhabwalaSverdlovKorotkovLikharev2004,
2DCIP-KinkhabwalaSverdlovKorotkovLikharev2004}. Briefly, the
conductor is ``fully frustrated" in the sense that the localized
sites are randomly and uniformly distributed, with a constant
``seed" density of states $\nu_0$, over both the rectangular 2D
sample of area $L \times W$ and a broad interval of ``seed" energies
$\varepsilon^{\left( 0 \right)}$. The full energy $U$ of the system
is the sum of the ``seed" energies of all occupied sites and the
electrostatic energy of the Coulomb interaction of the hopping
electrons with each other and the external capacitance:
\begin{equation}
U = \sum_{j} n_j\varepsilon^{\left( 0 \right)}_j + \frac{e^2}{2
\kappa} \sum_{j, k\neq j}\left(n_j-\frac{1}{2}\right)
\left(n_{k}-\frac{1}{2}\right) G\left( {\bf r}_j, {\bf r}_{k}
\right) + \frac{Q^2}{2C}. \label{eq:totalsystemenergy}
\end{equation}
Here $n_j$ (equal to either 0 or 1) is the occupation number of the
$j^{\textrm{th}}$ localized site, while $\kappa$ is the dielectric
constant of the insulating environment \cite{neutrality}. For the
simplest geometry of a 2D conductor connecting two semi-space-shaped
electrodes, the Green's function $G$ in
Eq.~(\ref{eq:totalsystemenergy}) may be simply expressed as a sum
over the infinite set of image charges in the electrodes:
\begin{eqnarray}
&G\left( {\bf r}_j, {\bf r}_k\right)& = \sum_{n=
-\infty}^{\infty}\left[ \frac{1}{\sqrt{\left( 2nL+x_k-x_j\right)^2 +
\left(
y_k-y_j\right)^2}} \right.\nonumber \\
&& \left. \,\,\,\,\,\,\,\,\,\, \,\,\,\,\,\,\,\,\,\,
\,\,\,\,\,\,\,\,\,\, \,\,\,\,\,\,\,\,\,\, -\frac{1}{\sqrt{\left(
2nL+x_k+x_j\right)^2 + \left( y_k-y_j\right)^2}} \right].
\label{eq:greenfunction}
\end{eqnarray}
For this geometrical model, the total charge $Q$ of the capacitor
(including the polarization component) is
\begin{equation}
Q = Q_i - \left[ N_e e + \sum_{j} e \left(n_j-\frac{1}{2}\right)
\frac{x_{j}}{L} \right], \label{eq:charge}
\end{equation}
where $Q_i$ is the initial charge and $x_j$ is the $j^{\textrm{th}}$
site position along the sample length $L$, while $N_e$ is the total
number of electrons that have passed through the conductor from the
start of the relaxation process until the given moment. In the limit
of large charge ($\left\vert Q \right\vert \gg Q_{R}$) the effect of
capacitance on hopping transport is equivalent to that of the
electric field $E=Q/CL$.

Electron hops are permitted from any occupied site $j$ to any
unoccupied site $k$ with the rate
\begin{equation}
\gamma_{jk}=\Gamma_{jk}\exp \left(-\frac{r_{jk}}{a}\right),
\label{eq:distancerateequation}
\end{equation}
where $a$ is the localization length, and
\begin{equation}
\hbar \Gamma_{jk}\left( \Delta U_{jk} \right)=g\frac{\Delta
U_{jk}}{1-\exp\left( -\Delta U_{jk}/k_{B}T\right) }.
\label{eq:energyrateequation}
\end{equation}
Here $\Delta U_{jk}$ is the difference of the total system energy
$U$ before and after the hop, and $g$ is a small dimensionless
parameter which affects only the scale of hopping conductivity
$\sigma_0\equiv g e^2/\hbar$. The numerical study has been carried
out using the classical Monte Carlo technique by Bortz, Kalos and
Leibowitz \cite{BKL} in the form suggested by Bakhvalov $\it
{et~al.}$ \cite{Bakhvalovetal1989}, which has become the de facto
standard for simulations of single-electron tunneling
\cite{Wasshuber}. An important feature of this algorithm is that it
is not slowed down by the gradual reduction of hopping rates at
charge relaxation.

\section{\label{sec:level1} Charge Relaxation Dynamics}

\begin{figure}
\begin{center}
\includegraphics[height=6.0in]{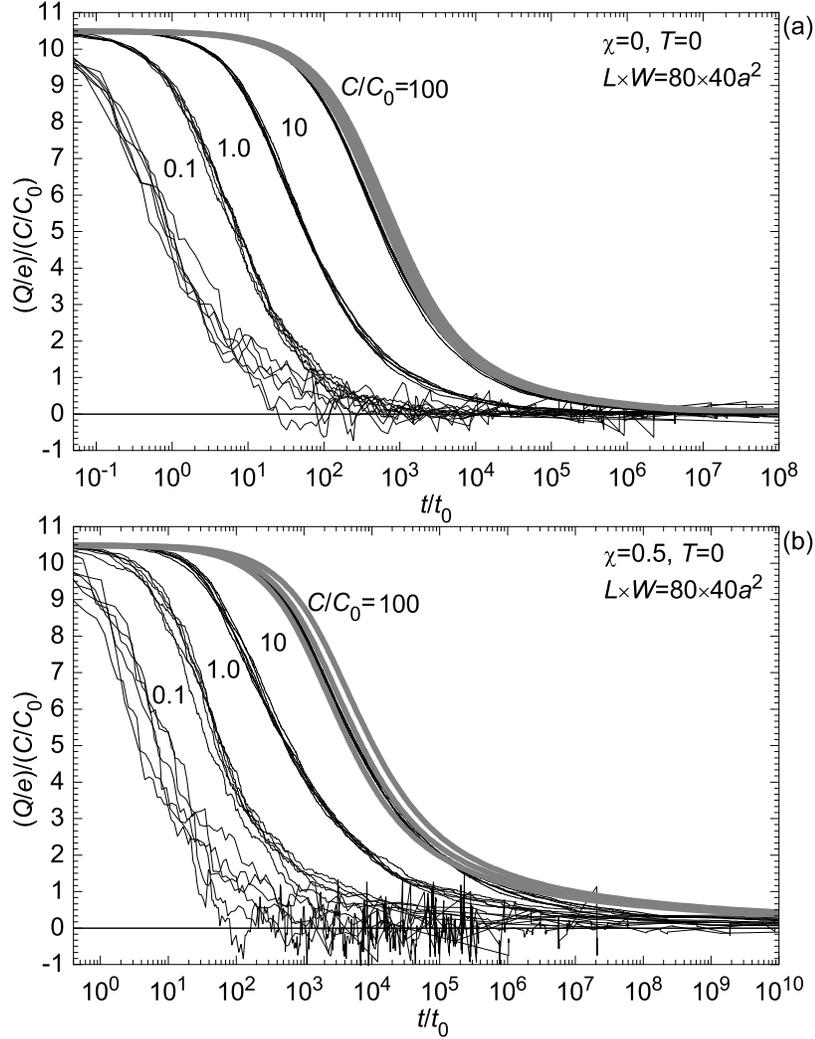}
\end{center}
\caption{Capacitance charge $Q$ relaxation (at $T=0$) for the cases
of (a) negligible ($\chi=0$) and (b) substantial ($\chi=0.5$)
Coulomb interaction of hopping electrons. Thin lines show Monte
Carlo results (for 6 realizations of each case) for several values
of external capacitance $C$, with fixed conductor size $L \times W =
80 \times 40 a^2$.  The thick gray curves correspond to the results
of the solution of Eq.~(\ref{eq:timederivativeofQres}) with
Eq.~(\ref{eq:zerocoulombhighelectricfieldconductivity}) for panel
(a) and Eq.~(\ref{eq:nonzerocoulombhighelectricfieldconductivity})
for panel (b) for $C/C_0 = 100$, with the central curve
corresponding to the best-fit parameters $A$ and $B$ and the outer
curves corresponding to the uncertainty in these parameters. (See
the text.) Time is measured in units of $t_0 \equiv \hbar \nu_{0}
a^2/g$, while capacitance is expressed in units of $C_{0} \equiv
e^{2}\nu_{0}a^2$.} \label{fig:zerotemperatureQRestimedependence}
\end{figure}

Figure~\ref{fig:zerotemperatureQRestimedependence} shows, by thin
lines, typical results of our Monte Carlo simulations for two values
of the dimensionless parameter of the Coulomb interaction strength,
$\chi \equiv e^{2} \nu_{0} a/ \kappa$. Note the logarithmic time
scale and the linear scale of $Q$; in such coordinates the
exponential relaxation of average charge in an $RC$ circuit with a
linear Ohmic resistor looks like a sharp step down at $t \approx
RC$. We indeed observe such behavior at hopping when the initial
electric field is low, i.e. in the high temperature limit. However,
motivated by prospects of practical applications
\cite{Likharev1999}, our main focus is on the opposite,
``high-field" (low-temperature) limit.
Figure~\ref{fig:zerotemperatureQRestimedependence} shows that in
this case the dynamics of discharge through the hopping conductor is
rather different: it slows down dramatically at $Q \rightarrow 0$.
This is exactly what should be expected from the previous studies of
variable-range hopping at constant applied field, which show that
the hopping conductance drops exponentially as the field decreases
\cite{MottBook, ShklovskiiBook, EfrosPollackCollection,
2DSverdlovKorotkovLikharev2001,
2DCLP-KinkhabwalaSverdlovKorotkovLikharev2004,
2DCIP-KinkhabwalaSverdlovKorotkovLikharev2004}. A qualitatively
similar dynamics is also typical for the qualitatively close (but
quantitatively different) problem of intrinsic relaxation in
electron glasses - see, e.g., recent publications~\cite{glass1,
glass2, glass3} and prior work cited therein.

It has turned out that most of the relaxation process, while the
charge is sufficiently large ($\left\vert Q \right\vert \gg Q_R$),
may be well described by the mean-field equation
\begin{equation}
\frac{dQ}{dt} = -I\left( T, E, \chi \right) = -\sigma\left(T, E,
\chi \right) E W, \label{eq:timederivativeofQres}
\end{equation}
where $\sigma\left(T, E, \chi \right)$ is the nonlinear conductance
in the constant applied field $E$. In the low-temperature limit
($k_{B}T \ll e E r$, where $r$ is the average length of the hops
contributing substantially into the current), we can use the
following analytical expressions obtained by fitting the results of
our numerical simulations of constant-field hopping within the same
model \cite{2DCLP-KinkhabwalaSverdlovKorotkovLikharev2004,
2DCIP-KinkhabwalaSverdlovKorotkovLikharev2004}:

(i) If Coulomb interaction is negligible, $\chi^3 \ll E/E_0$,
\begin{equation}
\frac{\sigma}{\sigma_0} \approx A \left( E, 0 \right) \exp
\left[-\left( B \left( E, 0 \right) \frac{E_0}{E} \right)^{1/3}
\right], \label{eq:zerocoulombhighelectricfieldconductivity}
\end{equation}
where $eE_{0}a \equiv 1/\nu_{0}a^2$, while $A \left( E, \chi
\right)$ and $B\left( E, \chi \right)$ are dimensionless, weak
functions of the applied field $E$ and Coulomb interaction strength
$\chi$. In a prior study
\cite{2DCLP-KinkhabwalaSverdlovKorotkovLikharev2004}, we have found
the best fit for the pre-exponential (model-specific) function to be
$A\left(E, 0\right) = \left(9.2 \pm 0.6
\right)\left(E/E_0\right)^{\left( 0.80 \pm 0.02\right)}$, with $B$
treated as a constant: $B\left( E, 0 \right) = 0.65\pm 0.02$.

(ii) If Coulomb effects are substantial, then
\begin{equation}
\frac{\sigma}{\sigma_0} \approx A \left( E, \chi \right) \exp
\left[-\left( B \left( E, \chi \right) \frac{\chi E_0}{E}
\right)^{1/2} \right].
\label{eq:nonzerocoulombhighelectricfieldconductivity}
\end{equation}
For the particular value of $\chi = 0.5$, a similar approach to
fitting gives \cite{2DCIP-KinkhabwalaSverdlovKorotkovLikharev2004}
$A \left( E , 0.5 \right) = \left(3.0 \pm 0.4
\right)\left(E/E_0\right)^{\left( 0.72 \pm 0.07\right)}$ with
$B\left( E, 0.5 \right) = 1.68\pm 0.07$.

For relatively low fields, $E \ll E_0$, these formulas describe the
so-called ``high-field" variable range hopping \cite{Shklovskii1973,
ApsleyHughes19741975, PollackRiess1976, RentzschShlimakBerger1979,
vanderMeerSchuchardtKeiper1982}.

Broad gray curves in
Fig.~\ref{fig:zerotemperatureQRestimedependence} show the results of
integration of the mean field equation using these formulas for one
value of capacitance $C/C_0 = 100$. (The middle curves correspond to
the best fit values, while the outer curves reflect the fitting
uncertainties specified above.) One can see that at $\left\vert Q
\right\vert \gg Q_{R}$ the relaxation results may be well described
by the mean-field approach. However, this approach does not work at
$Q \rightarrow 0$ where it predicts the complete relaxation of
charge, while in reality (and numerical experiment) the process
stalls at a certain ``residual" charge.

\section{\label{sec:level1} Residual Charge Statistics}

\begin{figure}
\begin{center}
\includegraphics[height=3.6in]{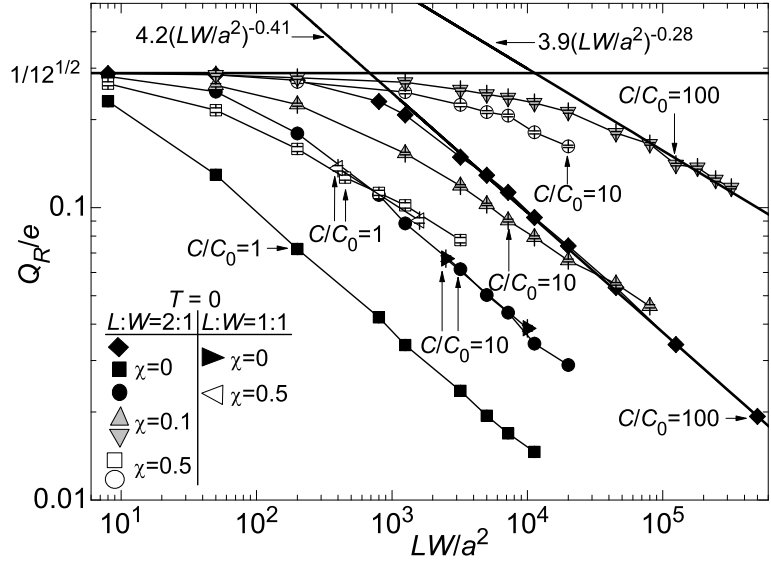}
\end{center}
\caption{The r.m.s$.$ value $Q_R$ of the residual charge at $T=0$
for negligible ($\chi=0$) and finite ($\chi$ = 0.1 and 0.5) Coulomb
interaction, as a function of the conductor area ($L \times W$) for
different external capacitances $C$, and two different aspect ratios
($L:W$ = 2:1 and 1:1). Each point represents data averaged over a
large number ($10^3$) of conductor samples with vertical error bars
corresponding to the uncertainty of such averaging. (Error bars are
shown on figure, unless smaller than the symbol size). Thin lines
are only guides for the eye. The bold horizontal line corresponds to
Eq.~(\ref{eq:QrmsUniformDistribution}), while the bold tilted lines
are the best power-law fits for large-sample data.}
\label{fig:QrmsVSlength}
\end{figure}

Figure~\ref{fig:QrmsVSlength} shows some of our results for the
r.m.s$.$ value $Q_R$ of the residual charge, obtained for a broad
range of ``macroscopic" parameters of the system, including external
capacitance $C$ and normalized Coulomb interaction strength $\chi$,
as a function of the conductor area $L \times W$. (These results do
not change noticeably if the systems are annealed after the
relaxation.)

For sufficiently small samples, the number of localized sites is so
low that no internal hopping events may occur within the energy
interval of interest, and the initial charge can only relax by
direct tunneling between the electrodes, giving changes of $Q$ in
multiples of $e$. The Coulomb blockade theory (see, e.g.,
Ref$.$~\cite{AverinLikharev1991}) shows that at low temperatures
such tunneling is blocked at $\left\vert Q \right\vert < e/2$. If
the initial charge $Q_i$ is random (as has been accepted in our
calculations), then the residual charge is uniformly distributed
within the range from $-e/2$ to $+e/2$, and the r.m.s$.$ residual
charge is
\begin{equation}
\frac{Q_{R}}{e} = \frac{1}{e} \left[ \int_{-e/2}^{e/2} Q^2
\frac{dQ}{e}\right]^{1/2} = \frac{1}{\sqrt{12}},
\label{eq:QrmsUniformDistribution}
\end{equation}
in a good accordance with the simulation results
(Fig.~\ref{fig:QrmsVSlength}).

On the other hand, if the conductor area is increased, $Q_{R}$
decreases, since there are more and more internal localized sites
available for further charge relaxation. Our results
(Fig.~\ref{fig:QrmsVSlength}) show also that $Q_{R}$ always
increases with capacitance $C$ and, at substantial Coulomb
interaction, with its strength $\chi$.

\begin{figure}
\begin{center}
\includegraphics[height=3.6in]{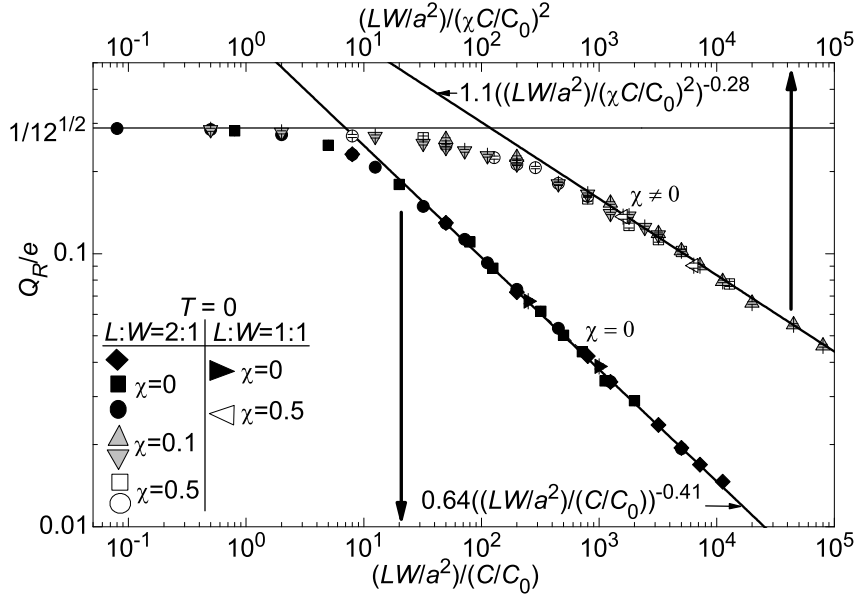}
\end{center}
\caption{The same results for $Q_{R}$ as in
Fig.~\ref{fig:QrmsVSlength}, re-plotted to emphasize their universal
scaling with system parameters. Solid lines show the best fits to
the asymptotic behavior of $Q_{R}$ for large samples.}
\label{fig:QrmsScaling}
\end{figure}

Rather unexpectedly, we have found that for a broad range of system
parameters, all these dependencies may be very well approximated by
``universal" laws, different for the cases when Coulomb interaction
is negligible ($\chi^3 \ll Q_R/CLE_0)$ or substantial - see
Fig.~\ref{fig:QrmsScaling}. In the former case, $Q_{R}/e =
F_{0}(X_0)$, where
\begin{equation}
X_{0} = \frac{L W}{a^2} \frac{C_0}{C} = L W \nu_{0} \frac {e^2}{C},
\label{eq:xxxhighelectricfieldQresoftime}
\end{equation}
while in the latter case $Q_{R}/e = F_{\chi}(X_\chi)$, where
\begin{equation}
X_{\chi} = \frac{L W}{a^2} \frac{C_0^2}{ \chi^2 C^2 } = \frac {L W
\kappa^{2}}{C^2}. \label{eq:xxxhighelectricfieldQresoftime}
\end{equation}
At small values of their arguments, both functions $F$ tend to
$1/\sqrt{12}$, in agreement with
Eq.~(\ref{eq:QrmsUniformDistribution}). Their asymptotic behavior is
also functionally similar, $F(X) \rightarrow DX^{-\beta}$ at $X
\rightarrow \infty$, but with different best-fit values of the
coefficients: for $\chi = 0$, $D = 0.64 \pm 0.01$ and $\beta = 0.41
\pm 0.01$, while for $\chi \sim 1$, $D = 1.1 \pm 0.1$ and $\beta =
0.28$, with the error bar about $0.03$ for the dependence on $C$ and
of the order of 0.01 for other variables contributing to $X_{\chi}$.

\section{\label{sec:level1} Discussion}

For the case of negligible Coulomb interaction, the asymptotic power
law for function $F_0(X_0)$ may be readily explained , using the
basic ideas of the Coulomb blockade \cite{AverinLikharev1991}.
Charge relaxation continues with the reduction of the system energy
(on the average, dominated by the capacitor energy $U$) until the
number $N$ of localized sites available for hopping becomes less
than one. If the capacitance charge before a hop is $Q$, the range
of capacitive energy of available initial sites is $\Delta U \sim
Q^{2}/2C$, so that the average number of such sites per unit area is
$n_i \sim \nu_{0} \Delta U \sim \nu_{0} Q^{2} /2C $, and their total
number in the sample of area $L \times W$ is $N_i \sim L W n_i \sim
L W \nu_0 Q^2/2C$. In order to estimate $N$, we need to multiply
$N_i$ by the average number $N_f$ of available final sites for each
initial site. For small changes of charge, $\vert \Delta Q \vert \ll
e$, the area $\vert \Delta x \vert \times W$ where such states can
reside is much smaller than the sample area $L \times W$, because
such charge change corresponds to a hop by distance $\vert \Delta x
\vert = L \times \vert \Delta Q \vert /e \ll L$. Hence $N_f \sim LW
\nu_0 (\vert \Delta Q\ \vert /e) (Q - \Delta Q)^2/2C$ and we get the
following estimate
\begin{equation}
N \sim N_i N_f \sim \left ( \frac {L W \nu_{0}}{2C} \right )^2 \frac
{Q^2 \vert \Delta Q \vert (Q - \Delta Q)^2}{e}.
\label{eq:N0}
\end{equation}
Now, from the natural requirement that $N$ drops below 1 as soon as
$\vert Q \vert$, $\vert \Delta Q \vert$, and $\vert Q-\Delta Q
\vert$ all become, on the average, of the order of $Q_R$, we get
\begin{equation}
\frac {Q_{R}}{e} \sim \left( \frac{L W \nu_{0} e^2}{C}
\right)^{-2/5} = {X_0}^{-2/5}, \label{eq:scaling0}
\end{equation}
which when compared to the power law $F(X)$ discussed above gives
$\beta = 2/5 = 0.40$, i.e$.$ inside the narrow interval $0.41 \pm
0.01$ given by the numerical experiment.

For the case of substantial Coulomb interaction of hopping
electrons, the situation is more complex - see, e.g$.$, the
discussion on pp$.$ 435-443 of Ref.~\cite{EfrosPollackCollection}.
It is well documented that ``external" transport (bringing electrons
into and out of the hopping conductor) may be well understood in
terms of the simple quasiparticles introduced by Efros and
Shklovskii \cite{ShklovskiiBook}, with energy
\begin{equation}
\varepsilon_j \equiv \varepsilon^{(0)}_j + \frac{e^2}{\kappa}
\sum_{l\neq j} \left( n_l-\frac{1}{2}\right) G\left( {\bf r}_j, {\bf
r}_l\right). \label{eq:singleparticleenergy}
\end{equation}
In 2D systems, their density of states at low energies $\epsilon$ is
given by the famous Coulomb-gap expression \cite{ShklovskiiBook}
\begin{equation}
\nu\left(\varepsilon \right) \approx \frac{2 \kappa^2}{\pi e^4}
\vert \varepsilon \vert. \label{eq:singleparticleDoS}
\end{equation}
If we naively  repeat the above calculation of $Q_R$, just replacing
$\nu_0$ with $\nu (\epsilon)$ from the last expression, we get
\begin{equation}
\frac {Q_{R}}{e} \sim \left( \frac{LW \kappa^2}{C^2} \right)^{-2/9}
= {X_{\chi}}^{-2/9},\label{eq:scalingchi}
\end{equation}
i.e$.$ the experimentally observed universality $(X_{\chi} = LW
\kappa^2 /C^2)$, but with an exponent $\beta = 2/9 \approx 0.22$
which is significantly outside of the experimental interval $0.28
\pm 0.01$.

Actually, for intra-sample transport, more adequate quasiparticles
may be the so-called ``dipole excitations" (essentially,
electron-hole pairs with correlated energies) whose density
$F(\Omega,r)$ depends on both the pair energy $\Omega$ and the
distance $r$ between the pair components (see
\cite{EfrosPollackCollection} p.435). In contrast to constant-field
transport, the residual charge statistics are dominated by
large-size pairs (hops), with $x-$component of the order of $L(\vert
\Delta Q \vert/e)$ and $y-$component of the order of $W$. If we
neglect, for such hops, the interaction of the pair components in
comparison with $\Omega$, then $F$ depends only on energy:
\begin{equation}
F(\Omega)=\int_{0}^{A} d\varepsilon_1 \int_{-A}^{0} d\varepsilon_2
\nu(\varepsilon_1) \nu(\varepsilon_2)
\delta(\varepsilon_1-\varepsilon_2-\Omega).
\end{equation}
For energies $\Omega$ much less than both the cutoff energy $A$ and
the Coulomb gap width, this integral yields
\begin{equation}
F=\left ( \frac{2 \kappa^2}{\pi e^4}\right )^2 \frac{\Omega^3}{6}.
\end{equation}
Now, following the arguments used above, we can accept $\Omega \sim
Q^2/2C$, and take $LW$ for the possible area of the pair centers,
and $L(\vert \Delta Q \vert/e)W$ for the pair area. After the
integration of $F$ from 0 to $\Omega$, for the possible number of
pairs within our energy range we get
\begin{equation}
N \sim \frac {1}{24} \left ( \frac{2 \kappa^2}{\pi e^4}\right )^2
\left (\frac {Q^2}{2C} \right )^4 L^2W^2 \frac {\vert \Delta Q
\vert}{e}.
\end{equation}
Again, requiring that $N \sim 1$ at $Q, \vert \Delta Q \vert \sim
Q_R$, we get back to the estimate given by Eq.
(\ref{eq:scalingchi}).

It is not quite clear presently whether the discrepancy between
these analytical arguments and the results of our numerical
experiments may be overcome by an account of electron-hole pairs of
smaller size, with strongly interacting pair components.

\section{\label{sec:level1} Offset Charge Grounding}

\begin{figure}
\begin{center}
\includegraphics[height=2.8in]{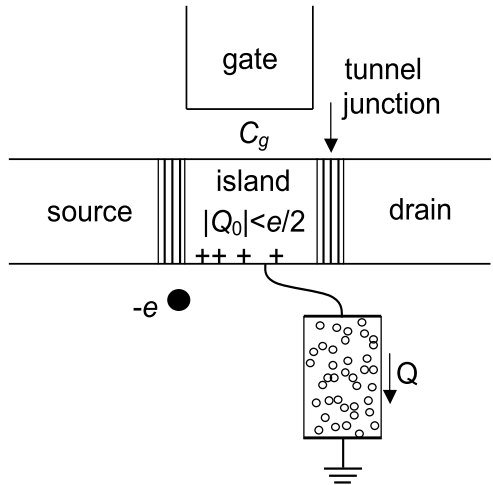}
\end{center}
\caption{Background charge ``grounding" using a hopping conductor
(schematically).} \label{fig:Ground}
\end{figure}

The results of this work allow one to estimate the prospects of
applying hopping conductors as ``grounding" devices for the random
background charge in single-electron devices. Figure
\ref{fig:Ground} shows this idea on the example of a single-electron
transistor \cite{AverinLikharev1991,Likharev1999}. Charged
impurities, randomly located in the vicinity of the transistor's
single-electron island, induce on it a net polarization charge. The
``integer" ($e$-multiple) part of this ``background" charge is
automatically compensated by tunneling through the transistor's
tunnel junctions, but its fractional part $-e/2 < Q_0 < +e/2$ cannot
be compensated in this way. This random charge is equivalent to a
random shift $\Delta V_g = Q_0 /C_g$ of the gate voltage; such
shifts are one of the main obstacles on the way toward integrated
circuits using single-electron devices, because for most of them the
tolerable background charge range is as narrow as $\sim 0.1e$
\cite{Likharev1999}. The problem may be solved by connecting the
single-electron island to ``ground" through a hopping conductor
which would provide a slow relaxation of the background charge
\cite{Likharev1999}. (For digital applications, the characteristic
relaxation time has to be much longer than at least the circuit
clock cycle, and more preferably the full time of the calculation
performed by the circuit.)

For typical hopping conductors technologically compatible with
silicon technology (e.g., amorphous semiconductors and metal
oxides), the dielectric constant $\kappa$ is of the order of 10,
while the electron effective mass $m \sim 0.2 m_0$. This gives the
localization radius $a \sim \hbar^2 \kappa /m e^2 \sim 3$ nm and the
level splitting scale $e^2/\kappa a \approx m e^4/\kappa^2 \hbar^2
\sim 30$ meV. In order to stay on the dielectric side of the
metal-insulator transition, the average distance between the
localized sites should be above $\sim 4a$ \cite{MottBook}; for the
3D density of states $\nu_3$ this gives the condition $\nu_3
\lesssim 10^{19}$ eV$^{-1}$cm$^{-3}$. This condition is well
satisfied, e.g., for most species of amorphous silicon, where
$\nu_3$ at mid-bandgap is of the order of 10$^{16}$
eV$^{-1}$cm$^{-3}$ (see, e.g., Ref$.$~\cite{a-Si}). For thin films
of such material with thickness $t \sim a \sim 3$ nm, the 2D density
of states $\nu_0 \sim 3 \times 10^{9}$ eV$^{-1}$cm$^{-2}$. For these
parameters, the Coulomb interaction parameter $\chi$ is much smaller
than 1, and we can use Eq.~(\ref{eq:scaling0}) for estimates. Even
for the least demanding applications of single-electron devices, the
electron addition energy $e^2/C$ should be at least 30 $k_BT$
\cite{Likharev1999}, so that according to Eq.~(\ref{eq:scaling0}),
$X_0$ has to be above $\sim 300$.

Let us accept $L=W$ in order to minimize the conductors' self-
(``stray") capacitance $C_s$ (which, as we will show shortly, may
present a major problem) at fixed area $L \times W$.  For the usual
conditions of low-temperature experiments with single-electron
devices, $T \sim 0.1$ K, $C$ may be of the order of $10^{-14}$ F, so
that with our parameters $L$ should be above $\sim$ 30 nm. This is
less than the typical length ($\sim$ 1 $\mu$m) of the
single-electron island in such experiments, so that the grounding
idea may actually work \cite{PriorObserv}.

On the other hand, for the most important case of room-temperature
single-electron devices ($ T \approx 300$ K), the island capacitance
should be much less, $C < 10^{-18}$ F, so that the quasi-continuous
conduction is only possible at $L \gtrsim$ 15 $\mu$m. Stray
capacitance $C_s$ of such a conductor would be larger than $\sim
10^{-15}$ F, i.e. much larger than $C$, thus increasing the total
effective capacitance of the island well above the acceptable value.

To summarize, our calculations indicate that the fractional charge
grounding is possible, but practicable only for low-temperature
experiments rather than for room-temperature single-electron
devices. Fortunately, by now an alternative way to solve (or rather
circumvent) the random background charge problem in digital
nanoelectronics has been suggested. This approach is based on
reconfigurable hybrid CMOS-nanodevice digital circuits which may be
re-routed around ``bad" devices  - see, e.g.,
Ref$.$~\cite{Springer2005}. Recent calculations have shown that this
approach may provide defect tolerance up to $\sim$ 10\% in memory
circuits and $> 20$\% in logic circuits. This is much higher than
the estimated lower bound on the fraction
($\sim$0.1\%~\cite{Likharev1999}) of single-electron devices whose
threshold is substantially shifted by random background charges.

\ack{The authors would like to thank B. I. Shklovskii for numerous
illuminating discussions. Useful comments by A. Efros, T. Grenet, A.
N. Korotkov, A. M{\"o}bius, M. Pollak and V. A. Sverdlov are also
gratefully acknowledged. The work was supported in part by the
Engineering Physics Program of the Office of Basic Energy Sciences
at the U.S. Department of Energy, and by the Semiconductor Research
Corporation. We also acknowledge the use of the following
supercomputer resources: our group's cluster $Njal$ (purchase and
installation funded by U.S. DoD's DURINT program via AFOSR), Oak
Ridge National Laboratory's IBM SP computer $Eagle $ (funded by the
Department of Energy's Office of Science and Energy Efficiency
program), and also IBM SP system $Tempest$ at Maui High Performance
Computing Center and IBM SP system $Habu$ at NAVO Shared Resource
Center (computer time granted by DOD's High Performance Computing
Modernization Program).}


\Bibliography{99}
%Introduction
\bibitem{MottBook} N. F. Mott and J. H. Davies, \emph{Electronic Properties of Non-Crystalline Materials}, \emph{2nd Ed.} (Oxford Univ. Press, Oxford, 1979); N. F. Mott, \emph{Conduction in Non-Crystalline Materials}, \emph{2nd Ed.} (Clarendon Press, Oxford, 1993).
\bibitem{ShklovskiiBook} B. I. Shklovskii and A. L. Efros, \emph{Electronic Properties of Doped Semiconductors} (Springer, Berlin, 1984).
\bibitem{EfrosPollackCollection} A. L. Efros and B. I. Shklovskii, ``Coulomb Interaction in Systems with Localized States", in \emph{Hopping Transport in Solids}, edited by M. Pollak and B.
Shklovskii (North-Holland, Amsterdam, 1991).
\bibitem{KoganBook} Sh. Kogan, \emph{Electronic Noise and Fluctuations in Solids} (Cambridge University Press, Cambridge, 1996).
\bibitem{AverinLikharev1991} D. V. Averin and K. K. Likharev, ``Single-Electronics", in \emph{Mesoscopic Phenomena in Solids}, edited by B. Altshuler \textit{et al.} (Elsevier, Amsterdam, 1991), pp. 173-271; see specifically p. 257.
\bibitem{MatsuokaLikharev1998} K. A. Matsuoka and K. K. Likharev, Phys. Rev. B \textbf{57}, 15613 (1998).
\bibitem{Review2002} D. Kaplan, Y. Kinkhabwala, A. Korotkov, V. Sverdlov, and K. Likharev, ``Sub-electron Charge Transport in Nanostructures", in \emph{Proc. of the 20th Symposium on Energy Engineering Sciences}, ANL, Agronne, IL, 2002), pp.
231-240.
\bibitem{NavehAverinLikharev1998} Y. Naveh, D. Averin, and K. Likharev, Phys. Rev. B \textbf{58}, 15371 (1998).
\bibitem{Likharev1999} K. K. Likharev, Proc. of IEEE \textbf{87}, 606 (1999).
\bibitem{JongBeenakker1997} M. J. M. de Jong and C. W. J. Beenakker, ``Shot Noise in Mesoscopic Systems", in \emph{Mesoscopic Electron Transport}, edited by L. L. Sohn, L. P. Kouwenhoven, and G. Sch\"{o}n, NATO ASI Series Vol. 345 (Kluwer Academic Publishers, Dordrecht, 1997), p.225.
\bibitem{BlanterButtiker2000} Ya. M. Blanter and M. Buttiker, Phys. Repts. \textbf{336}, 2 (2000).
\bibitem{Kuznetsovetal2000} V. V. Kuznetsov, E. E. Mendez, X. Zuo, G. Snider, and E. Croke, Phys. Rev. Lett. \textbf{85}, 397 (2000).
\bibitem{Roshkoetal2002} S. H. Roshko, S. S. Safonov, A. K. Savchenko, W. R. Tribe, and E. H. Linfield, Physica E \textbf{12}, 861 (2002).
\bibitem{1DKorotkovLikharev2000} A. N. Korotkov and K. K. Likharev, Phys. Rev. B \textbf{61}, 15975 (2000).
\bibitem{2DSverdlovKorotkovLikharev2001} V. A. Sverdlov, A. N. Korotkov, and K. K. Likharev, Phys. Rev. B \textbf{63}, 081302(R) (2001).
\bibitem{2DCLP-KinkhabwalaSverdlovKorotkovLikharev2004} Y. A. Kinkhabwala, V. A. Sverdlov, A. N. Korotkov, and K. K. Likharev, J. Phys.: Condens.
Matter \textbf{18}, 1999 (2006).
\bibitem{2DCIP-KinkhabwalaSverdlovKorotkovLikharev2004} Y. A. Kinkhabwala, V. A. Sverdlov and K. K. Likharev, J. Phys.: Condens.
Matter. \textbf{18}, 2013 (2006).
\bibitem{Lambe} J. Lambe and R. C. Jaklevic, Phys. Rev. Lett. \textbf{22}, 1371 (1969).
\bibitem{Kuzmin} L. S. Kuzmin and K. K. Likharev, JETP Lett. \textbf{45}, 495 (1987).
\bibitem{neutrality} Following most studies of the Coulomb interaction at hopping, we keep the conductor electro-neutral by adding an effective
background charge of $-e/2$ to each localized site.
\bibitem{BKL} A. B. Bortz, M. H. Kalos, and J. L. Leibowitz, J.
Comp. Phys. \textbf{17}, 10 (1975).
\bibitem{Bakhvalovetal1989} N. S. Bakhvalov, G. S. Kazacha, K. K. Likharev, and S. I. Serdyukova, Sov. Phys. JETP \textbf{68}, 581 (1989).
\bibitem{Wasshuber} C.~Wasshuber, \emph{Computational Single-Electronics} (Springer, Berlin, 2001), Ch. 3.
\bibitem{glass1} C. J. Adkins, J. D. Benjamin, J. M. D. Thomas, J. W. Gardner, and A. J. McCeown, J. Phys. C \textbf{17}, 4633 (1984).
\bibitem{glass2} Z. Ovadyahu and M. Pollak, Phys. Rev. B \textbf{68}, 184204 (2003).
\bibitem{glass3} T. Grenet, Eur. Phys. J. B \textbf{32}, 275 (2003); Phys. Stat. Sol. (c) \textbf{1}, 9 (2004).
\bibitem{Shklovskii1973} B. I. Shklovskii, Sov. Phys. Semicond. \textbf{6}, 1964 (1973).
\bibitem{ApsleyHughes19741975} N. Apsley and H. P. Hughes, Philos. Mag. \textbf{30}, 963 (1974); \textbf{31}, 1327 (1975).
\bibitem{PollackRiess1976} M. Pollack and I. Riess, J. Phys. C \textbf{9}, 2339 (1976).
\bibitem{RentzschShlimakBerger1979} R. Rentzsch, I. S. Shlimak and H. Berger, Phys. Status Solidi A \textbf{54}, 487 (1979).
\bibitem{vanderMeerSchuchardtKeiper1982} M. van der Meer, R. Schuchardt and R. Keiper, Phys. Status Solidi B \textbf{110}, 571 (1982).
\bibitem{a-Si} T. Sameshita and S. Usui, J. Appl. Phys. \textbf{70}, 1281 (1991).
\bibitem{PriorObserv} Actually, the first qualitative observations of relaxation of sub-electron background
charge to $Q_R \ll e$ in early experiments \cite{Lambe,Kuzmin} may
be considered as the first, albeit unintentional implementations of
this idea.
\bibitem{Springer2005} K. K. Likharev and D. V. Strukov, ``CMOL: Devices, Circuits, and Architectures",
in \emph{Introducing Molecular Electronics}, edited by G. Cuniberti \textit{et al.} (Springer, Berlin, 2005), pp. 447-477.

\endbib
\end{document}